\documentclass[11pt,a4paper,dvips]{scrartcl}
\usepackage{cite,mcite}
\usepackage{graphicx}
\usepackage{subfig}
\usepackage{lcd}
\usepackage{lcd_title,ifthen}
\usepackage{mathptmx}
\usepackage{helvet}
\usepackage{amsmath}
\usepackage{color}
\usepackage{units}
\usepackage{url}
\usepackage{caption}
\usepackage[margin=1in]{geometry}

\makeatletter
\def\url@myurlfontstyle{%
  \@ifundefined{selectfont}{\def\UrlFont{\sf}}{\def\UrlFont{\small\ttfamily}}}
\makeatother
\long\def\symbolfootnote[#1]#2{\begingroup%
\def\thefootnote{\fnsymbol{footnote}}\footnote[#1]{#2}\endgroup}
 \lcdnote{2010-005}
\newlength{\capindent}
\newlength{\capwidth}
\newlength{\figwidth}
\newcommand{\icaption}[2][!*!,!]{\hspace*{\capindent}%
  \begin{minipage}{\capwidth}
    \ifthenelse{\equal{#1}{!*!,!}}%
      {\caption{#2}}%
      {\caption[#1]{#2}}
      \vspace*{3mm}
  \end{minipage}}
\def\susy#1{\ensuremath{\tilde{\mathrm{#1}}}}%

\def\smuon     #1{\ensuremath{\susy{\mu}^{#1}}}

\def\neutralino#1{\ensuremath{\susy{\chi}_{#1}^0}}

\def\EV30{\ensuremath{E_{\mathrm{v30}}}}%
\def\Ebar{\ensuremath{E\hspace{-.23cm}/\hspace{+.01cm}}}
\def\EM25{\ensuremath{\Ebar_{25}}}
\def\EMF25{\ensuremath{\Ebar^{\perp}_{25}}}
\def\ECM60{\ensuremath{\Ebar^b_{60}}}

\def\TKM25{\ensuremath{N_{tk}^{25}}}

\def\simge{\ \lower -2.5pt\hbox{$>$} \hskip-8pt \lower 2.5pt \hbox{$\sim$}\ }
\def\simle{\ \lower -2.5pt\hbox{$<$} \hskip-8pt \lower 2.5pt \hbox{$\sim$}\ }

\begin{document}
\begin{titlepage}
%
\vskip 35mm
%
\mydocversion
%
\title{Physics requirements for Scalar Muons searches at CLIC}

\author{M. Battaglia\affiliated{1} \affiliated{2}, J-J. Blaising\affiliated{3}
             }
\affiliations{\affiliation[1]{CERN, Geneva, Switzerland},
\affiliation[2]{University of California at Santa Cruz, Santa Cruz, CA, USA}, \newline
\affiliation[3]{Laboratoire d'Annecy-le-Vieux de Physique des Particules, Annecy-le-Vieux, France}}
%
\date{\today}
%
\begin{abstract}
\noindent
The determination of smuon and neutralino masses in smuon pair production is an important part 
of the program of spectroscopic studies of Supersymmetry at a high energy linear collider.
In this note we report the first results of a study of $e^+e^- \to \tilde \mu_R^+ \tilde \mu_R^-$ 
in a high-mass, cosmology-motivated Supersymmetric scenario at 3~TeV at CLIC. This process is a 
good example to study requirements on the beam energy spectrum and polarisation and the track 
momentum resolution in a simple final state. 
We discuss the expected accuracy on the mass measurements as a function of the momentum resolution, 
luminosity spectrum, beam polarisation and time stamping capability. 
Results obtained at generator level are validated by comparison to full simulation and reconstruction. 
Preliminary requirements for the detector performances and beam polarisation are presented.
\end{abstract}

\end{titlepage}
%
%
\section{Introduction}
One of the main objectives of linear collider experiments is the precision spectroscopy 
of new particles predicted in theories of physics beyond the Standard Model (SM), such as  
Supersymmetry (SUSY). Since some, or most, of these particles may have masses of 
$\cal{O}\mathrm{(1~TeV)}$, these studies may be central to the physics program of a 
multi-TeV $e^+e^-$ linear collider, such as CLIC.

In this note, we study the production of the supersymmetric partners of the muon in a 
specific SUSY scenario, where we assume R--parity conservation within the so-called 
constrained Minimal Supersymmetric extension of the SM (cMSSM).
In this model the neutralino (\neutralino{1}) is the lightest supersymmetric particle and the specific 
parameters of the benchmark point~\cite{Battaglia:2003ab} are chosen to make it compatible 
with current collider and cosmology data. In particular, the properties of the lightest neutralino
are such that it generates the correct amount of relic dark matter density in the universe, as obtained
from the analysis of the WMAP data~\cite{Komatsu:2008hk}.
Scalar muons ($\smuon{\pm}_R$ and $\smuon{\pm}_L$) are the supersymmetric partners of the right- and 
left-handed charged muons. Smuons are produced in pair through $s$-channel $\gamma/\mathrm{Z}$ exchange 
in the process $e^+e^- \to \tilde \mu_R^+ \tilde \mu_R^-$ and each decay into an ordinary muon and a 
neutralino, \neutralino{1}. The neutralino, being weakly-interacting, escapes detection.
Therefore, the experimental signature of the process is two oppositely charged muons plus missing energy. 
This study concentrates on the lightest smuon, $\smuon{\pm}_R$, which, for the chosen model parameters, 
has a mass of 1108.8~GeV , while the mass of the lightest neutralino is 554.3~GeV.
The accurate determination of their masses is an essential part of the spectroscopy study of a high 
mass SUSY scenario at CLIC. A study of the variation of the predicted relic dark mass density in the 
universe $\Omega h^2$ with the lightest neutralino mass in the cMSSM shows that a $\pm$1.0~GeV uncertainty 
on its mass corresponds to a $\pm$0.05 relative uncertainty on $\Omega h^2$, i.e.\ the current accuracy 
from cosmic microwave background observations~\cite{Komatsu:2010fb}.
The main aim of this study is to assess the requirements for a detector at CLIC operating at a 
centre-of-mass energy, $\sqrt{s}$, of 3~TeV as a function of the track 
momentum resolution, luminosity spectrum and beam polarisation. The reconstruction of the particle masses 
through the endpoints of the muon momentum spectrum is a good example for these requirements since the 
analysis is particularly simple and can be carried out using a simple momentum smearing on generator-level 
observables. Results are validated using full simulation and reconstruction with the CLIC-ILD detector model. 

\section{Simulation data sample} \label{dtsmcs}

The simulation is performed for the cMSSM parameters of point K' of ref.~\cite{Battaglia:2003ab}.
In the cMSSM the mass parameters are defined at the GUT scale. The subsequent evolution to the 
electro-weak scale is performed using the renormalisation group equations of 
{\sc Isasugra~7.69}~\cite{Paige:2003mg}. Signal events are generated using 
{\sc Pythia~6.125}~\cite{Sjostrand:2006za}. At 3 TeV, the production cross section for the 
process $\ee \rightarrow \tilde \mu_R^+ \tilde \mu_R^-$, for unpolarised beams, is 0.71~fb.
Beamstrahlung effects on the luminosity spectrum are included using results of the CLIC beam 
simulation for the 2008 accelerator parameters~\cite{Braun:2008zzb}. 
Initial state radiation (ISR) is included in the event generation in {\sc Pythia}.
The following background processes have been included in the background calculation:

\begin{table} [h]
\begin{tabular}{ll} 
Process & Cross section \\
$\ee \rightarrow \mathrm{W^+ W^-} \rightarrow \mu^+ \mu^- \nu_{\mu} \nu_{\mu}$   &~~ $\mathrm \sigma$=10.5 fb \\
$\ee \rightarrow \mathrm{Z^0 Z^0} \rightarrow \mu^+ \mu^- \nu \nu$               &~~ $\mathrm \sigma$=0.5 fb \\
$\ee \rightarrow \mathrm{Inclusive~SUSY} \rightarrow \mu^+ \mu^- X$              &~~ $\mathrm \sigma$=0.4 fb \\
$\ee \rightarrow \mu \nu_{e} \mu \nu_{e}$ (inclusive SM)                         &~~  $\mathrm \sigma$=135 fb \\
\end{tabular}
\end{table}

The first three processes have been simulated with {\sc Pythia}. 
In addition, the inclusive SM process  $e^+e^- \to \mu^+ \mu^- \nu_{e} \nu_{e}$ is 
generated using CompHep~\cite{Comphep}, removing the contributions from the $W^+W^-$ and $Z^0Z^0$ 
diagrams, to avoid double counting. The estimated cross section is 135~fb. In the background study 
we neglect the $e^+e^- \to \mu^+ \mu^- \nu_{\mu} \nu_{\mu}$ contribution not due to  $W^+W^-$ and 
$Z^0Z^0$ decays, since its cross section is only $\simeq$0.2~fb. We assume a data sample corresponding 
to an integrated luminosity of 2~ab$^{-1}$ taken at a nominal $\sqrt{s}$ energy of 3~TeV, corresponding 
to $\simeq$3.5 years (1 year = $10^{7}$~s) of run at the nominal CLIC luminosity of 
6$\times$10$^{34}$~cm$^{-2}$s$^{-1}$. Beam polarisation is in general extremely helpful in the 
study of SUSY processes both to improve the signal-to-background ratio and as an 
analyser~\cite{MoortgatPick:2005cw}. We consider here three options for beam polarisation: 
i) unpolarised beams, ii) P($e^-$)=+80~\% and P($e^+$)=0~\% and iii) P($e^-$) = +80~\% and 
P($e^+$)=-60~\%.  
The main benefits of beam polarisation for this analysis are the suppression of the $W^+W^-$ 
background (by a factor of five for ii, and ten for iii)), the enhancement of the smuon
cross section (by a factor 1.5 for ii), and 2.3 for iii)) and the possibility to disentangle 
$\tilde \mu_R \tilde \mu_R$ from $\tilde \mu_R \tilde \mu_L$ and $\tilde \mu_L \tilde \mu_L$ 
production. In this analysis we use observables at the generator level applying a track momentum 
smearing. Results are validated by comparing with fully simulated and reconstructed events in 
section 3.3.7.   
 
%
\section{Analysis Procedure}
\subsection{Signal topologies and event pre-selection}
\begin{figure}[h]
\begin{center}
\includegraphics[width=7.0cm,clip]{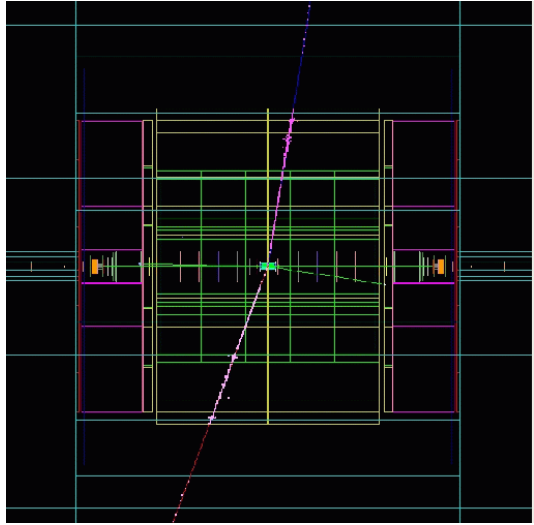}
\end{center}
  \caption{Display of a simulated  $e^+e^- \to \tilde \mu_R^+ \tilde \mu_R^- \to \mu^+ \mu^- \tilde \chi^0_1 \tilde \chi^0_1$}
  \label{fig:evt}
\end{figure}

\begin{figure}[h]
\begin{center}
\includegraphics[width=15.0cm,clip]{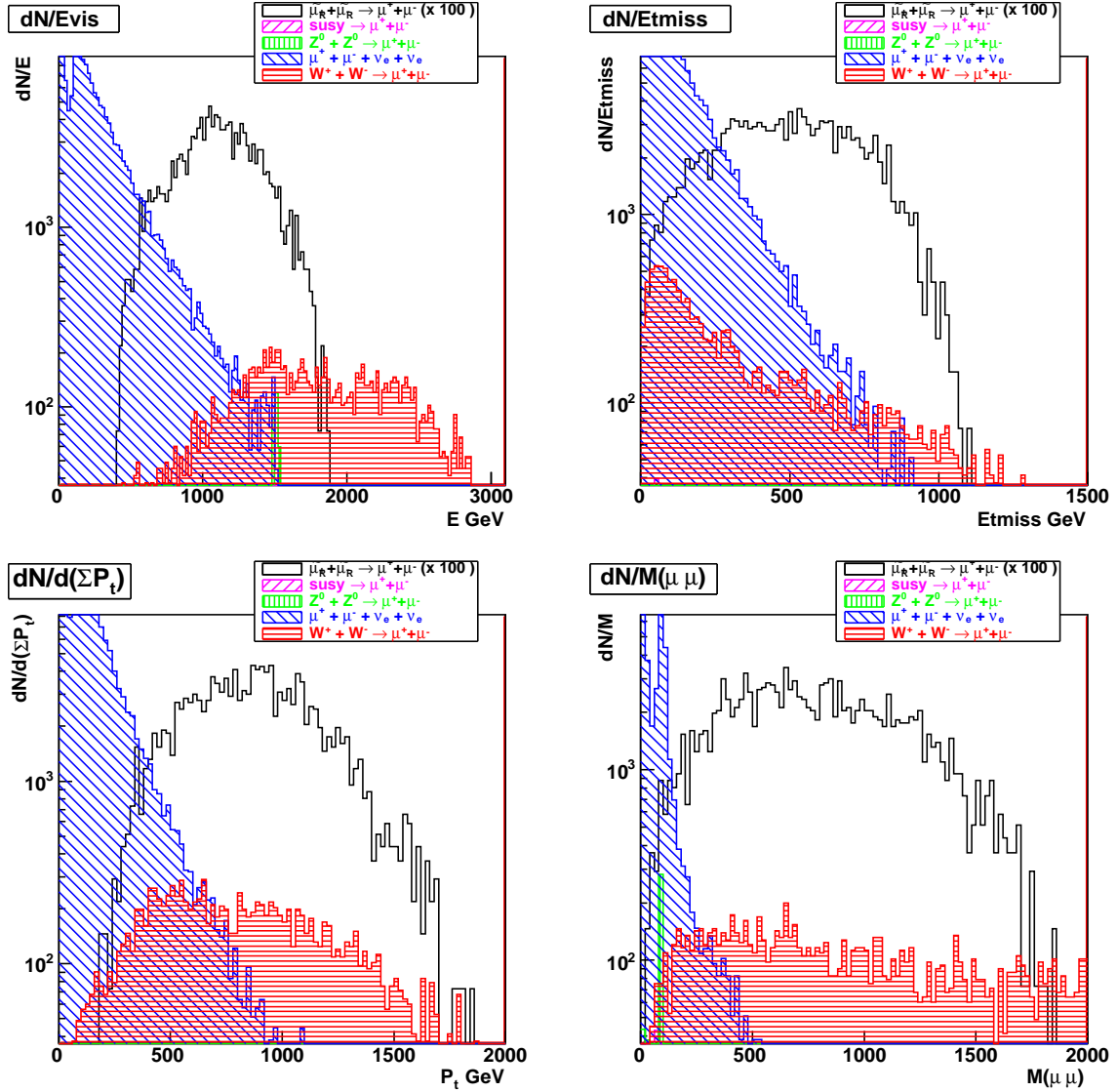}
\end{center}
  \caption{Discriminating variables used in the combined likelihood function:
   (upper left) $\mathrm E_{\rm vis}$ visible energy, 
   (upper right) $E_{\rm t miss}$ missing transverse energy,
   (lower left)  $\sum \rm p_{t}$ sum of the $\rm p_{t}$ of the muons
    and (lower right) $M_{( \mu \mu) }$ invariant mass of the two muons
   }
  \label{fig:c2mu-p1}
\end{figure}

\begin{figure}[h]
\begin{center}
\begin{tabular}{c}
\includegraphics[width=15.0cm,clip]{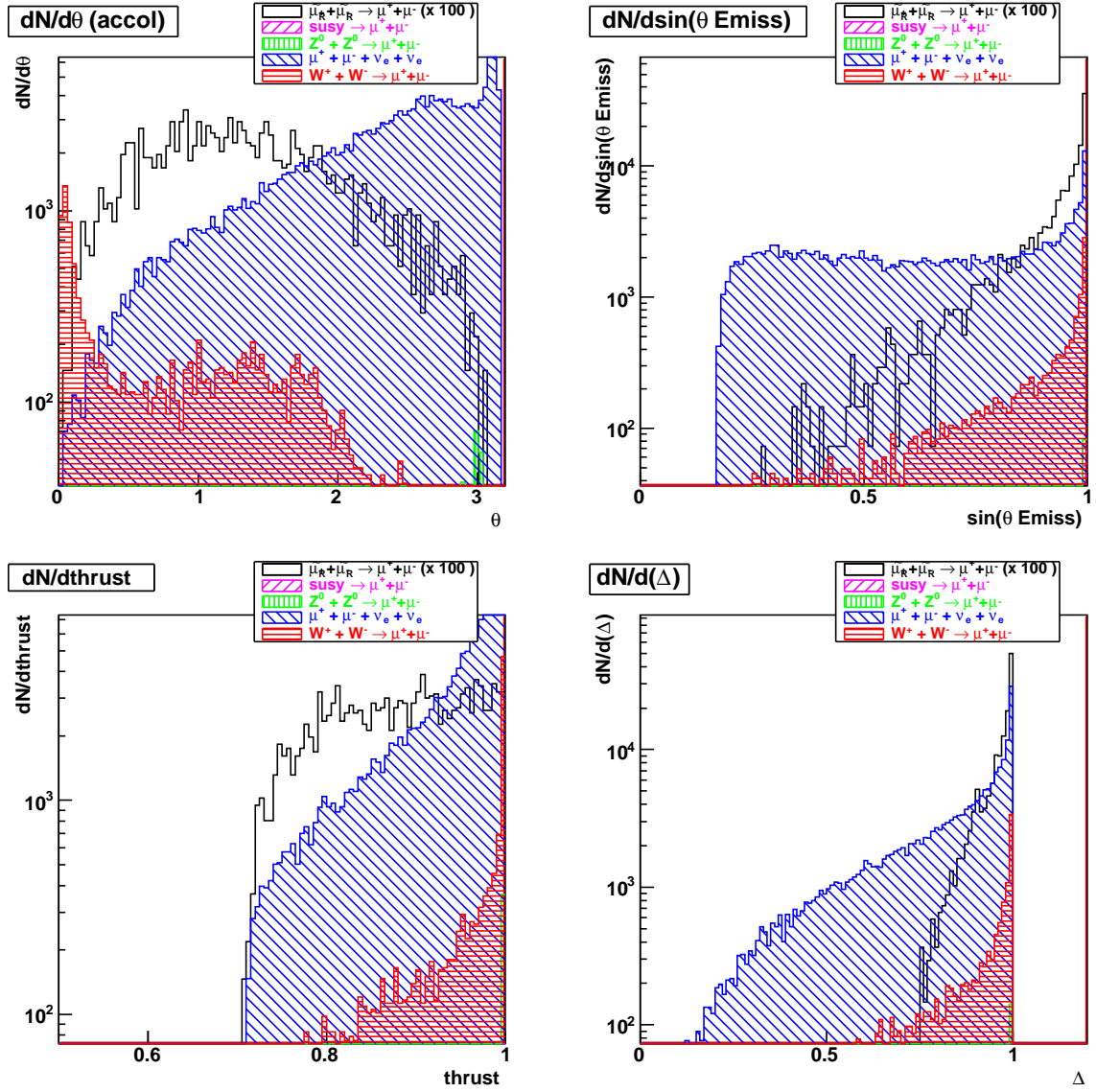}
\end{tabular}
\end{center}
  \caption{Discrimination variables used in the combined likelihood function:
  (upper left) $\theta$ acollinearity,
  (upper right) $\sin(\theta_{\rm Emiss}$ missing energy direction,
  (lower left)  thrust of the two muon system and
  (lower right) distribution of the variable $\Delta$ (see text) }
  \label{fig:c2mu-p2}
\end{figure}

The signal process has two undetected \neutralino{1}'s in the final state.
Therefore, the main characteristics of signal events are large
missing transverse momentum, missing energy and acoplanarity (see Figure~\ref{fig:evt}). 
Despite the striking signature of two muons and large missing energy, the small 
anticipated signal production cross section at the K' benchmark point, makes this 
analysis rather challenging. In our analysis, the signal selection proceeds as follows. 
First, we apply an event pre-selection, which requires two oppositely-charged 
muons with $p_{t} \ge$ 5~GeV and $|\cos \theta| < 0.985$, where $\theta$ is the 
particle polar angle. Next, we combine the values of the signal probabilities for 
the following discriminating variables into a global likelihood variable {\tt Prob}: 
\begin{itemize}
\item visible energy $E_{\rm vis}$,
\item missing transverse energy $E_{\perp miss}$,
\item sum of transverse momentum of the muons $\sum |p_{t}|$,
\item maximum acollinearity and acoplanarity,
\item polar angle of the missing energy vector ($\theta_{miss}$)
\item invariant mass of the two muons,
\item the thrust of the two muons,
\item unbalance of the muon momenta $\Delta$ 
\item missing mass $M_{\rm mis}$
\end{itemize}
where $\Delta= \left(1-\frac{ ( P_{\mu1}-P_{\mu2} )^{2} } {(  P_{\mu1}+P_{\mu2})^{2} } \right)^{1/2}$
and $M_{\rm mis}=(s+M_{\rm vis}^2-2\sqrt{s} E_{\rm vis})^{1/2}$
with the missing mass calculated from the visible energy $E_{\rm vis}$ and momentum $P_{\rm vis}$,
and $M_{\rm vis}=(E_{\rm vis}^2-P_{\rm vis}^2)^{1/2}$. Figures~\ref{fig:c2mu-p1} and~\ref{fig:c2mu-p2}
show the distributions of some of the discriminating observables for signal and 
background samples after pre-selection and requiring $\sqrt{s} \ge 2500$~GeV.

\newpage

\subsection{Final selection efficiency and background estimate}
\begin{figure}[h]
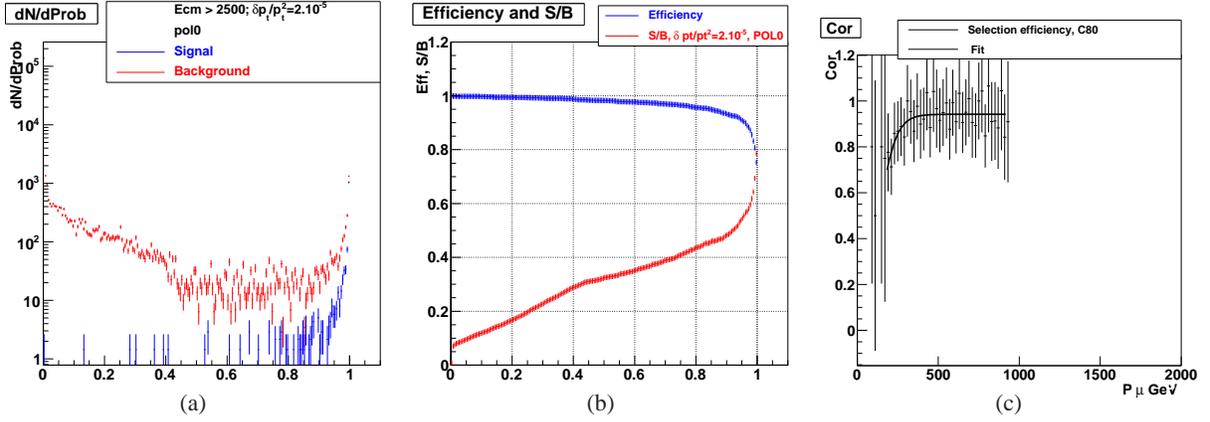

\begin{center}
\begin{tabular}{ccc}
 \subfloat[]{\includegraphics[width=5.0cm,clip]{fsel_SandBPROBpol0smear2e-5.epsi}} &
 \subfloat[]{\includegraphics[width=5.0cm,clip]{fsel_EFFIPROBpol0smear2e-5.epsi}} &
\subfloat[]{\includegraphics[width=5.0cm,clip]{fsel_psel_H1LPA205_smear2e-5_pol0_0BX_e2500C80_EFC.epsi}} \\
\end{tabular}
\end{center}
\caption{(a) (left panel) distribution of the combined probability variable for signal events (blue)
and background events (red); $\delta \rm p_{t}/p_{t}^2$ = $2 \times 10^{-5}$~GeV$^{-1}$, 
(b) (middle panel) efficiency and S/B as a function of the probability value without polarisation, 
(c) (right panel) selection efficiency for a probability cut larger than 0.8 as a function of muon momentum.}
\label{fig:h1prob1}
\end{figure}

The normalised signal-to-background ratio, S/B, values of these variables, as well as the 
combined probability {\tt Prob} are computed for different detector resolution assumptions: 
$\delta \rm p_{t}/p_{t}^{2}$= $2 \times 10^{-5}$, $4 \times 10^{-5}$, $6 \times 10^{-5}$, 
$8 \times 10^{-5}$ and $2 \times 10^{-4}$~GeV$^{-1}$. 
Fig.~\ref{fig:h1prob1} shows the 
distribution of the combined probability for signal and background events, the selection efficiency 
and the signal-over-background ratio as a function of the combined probability value, as well as
the signal selection efficiency as a function of muon momentum.

\begin{figure}[h]
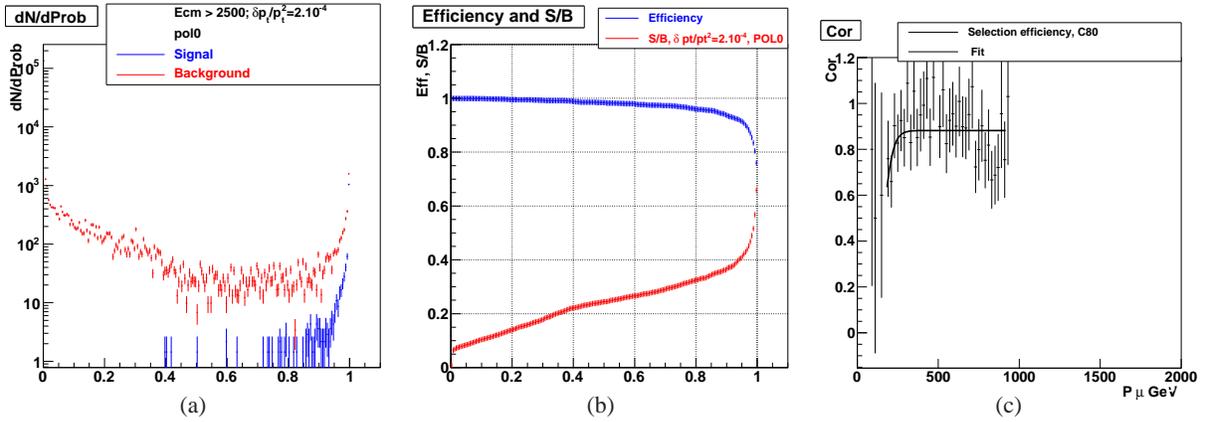

\begin{center}
\begin{tabular}{ccc}
 \subfloat[]{\includegraphics[width=5.0cm,clip]{fsel_SandBPROBpol0smear2e-4.epsi}} &
 \subfloat[]{\includegraphics[width=5.0cm,clip]{fsel_EFFIPROBpol0smear2e-4.epsi}} &
 \subfloat[]{\includegraphics[width=5.0cm,clip]{fsel_psel_H1LPA205_smear2e-4_pol0_0BX_e2500C80_EFC.epsi}} \\
\end{tabular}
\end{center}
\caption{Same as Fig,~\ref{fig:h1prob1} for $\delta \rm p_{t}/p_{t}^2$ = $2 \times 10^{-4}$~GeV$^{-1}$. 
In (c) the deformation of both the lower and the upper end of the spectrum after selection cuts is visible.}
\label{fig:h1prob2}
\end{figure}

There are two main effects on the muon momentum distribution in selected events. First, the efficiency of 
the selection on the combined probability is not flat with the muon momentum. Therefore, a cut on this 
variable introduces an inefficiency at the lower edge of the distribution and a subsequent bias towards 
higher momenta, see Figure~\ref{fig:h1prob1}(c). This inefficiency increases with the value of the 
probability cut applied. The inefficiency and the bias increase also when the momentum resolution 
degrades. Fig.~\ref{fig:h1prob2} shows the same distributions for 
$\delta \rm p_{t}/p_{t}^2$ = $2 \times 10^{-4}$~GeV$^{-1}$, Fig.~\ref{fig:h1prob2}(c) shows a
bias towards higher momenta. This effect is accounted and corrected for in the fits performed for 
signal+background (see Figure~\ref{fig:H1SPB1} (b))
Both beamstrahlung and momentum resolution introduce a smearing of the upper momentum edge. 
Both effect have a potential impact on the statistical accuracy and the bias in extracting 
the SUSY particle masses from a fit to the reconstructed momentum distribution, as discussed below.

\begin{figure}[h]
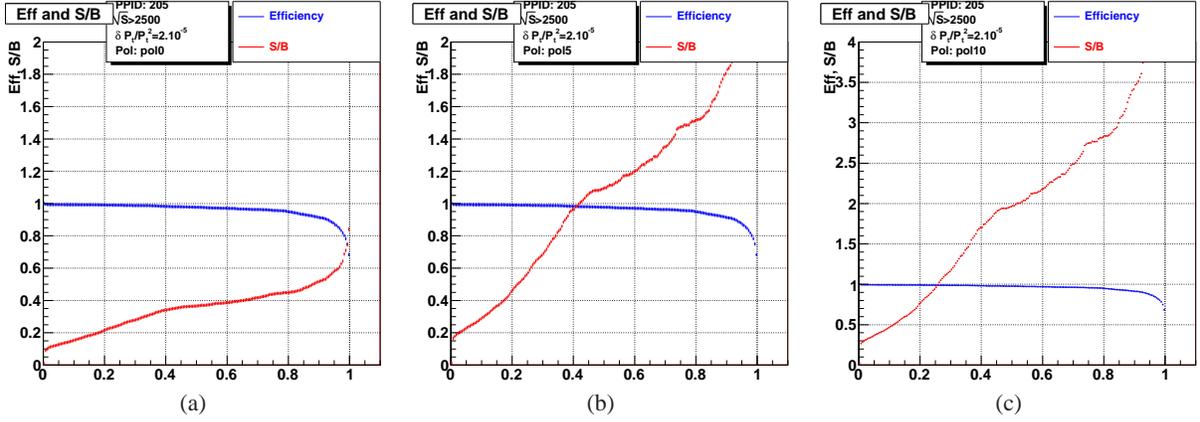

\begin{center}
\begin{tabular}{ccc}
 \subfloat[]{\includegraphics[width=5.0cm,clip]{fsel_205_smear2e-5_pol0_0BX_e2500_EFFI.epsi}} &
 \subfloat[]{\includegraphics[width=5.0cm,clip]{fsel_205_smear2e-5_pol5_0BX_e2500_EFFI.epsi}} &
 \subfloat[]{\includegraphics[width=5.0cm,clip]{fsel_205_smear2e-5_pol10_0BX_e2500_EFFI.epsi}} \\
\end{tabular}
\end{center}
\caption{efficiency and S/B as a function of the probability value for
different polarisation options, (a) no polarisation
(b) $80\%~\mathrm{e^{-}}$~polarisation~ and (c) $80\%~\mathrm{e^{-}~+~60\%~e^{+}}$ polarisation 
}
\label{fig:hpol}
\end{figure}

The Signal-over-background ratio depends also on the beam polarisation.
Fig.~\ref{fig:hpol} shows the efficiency and S/B as a function of the probability value for
different polarisation options.

Table~\ref{tab2} lists the number of signal (S) and background (B) events, the selection efficiency 
$\epsilon$ and the S/B ratio for different values of the probability cut, momentum resolution, 
polarisation and time stamping values.

\begin{table} [h]
\begin{center}
\begin{tabular}{|l|l|c|c|c|c|c|c|} \hline

 Prob cut & $\delta \rm {p_{t}} / {p_{t}}^{2}$ & pol & BX & $\mathrm N_{\rm sig}$ &  $\mathrm N_{\rm bkg}$
          & $\epsilon$ & $\mathrm N_{\rm sig} /  \mathrm N_{\rm bkg}$  \\  
      &                      & ($e^{-}/e^{+})$ &  &      &      &      &       \\ \hline
 0.80 & 0                    & 0/0             & 0 & 1315 & 2937 & 0.93 & 0.45  \\  \hline
 0.80 & 2 $\times$ 10$^{-5}$ & 0/0             & 0 & 1319 & 2984 & 0.93 & 0.44 \\  \hline
 0.80 & 4 $\times$ 10$^{-5}$ & 0/0             & 0 & 1319 & 2953 & 0.93 & 0.45 \\  \hline
 0.80 & 6 $\times$ 10$^{-5}$ & 0/0             & 0 & 1318 & 3098 & 0.93 & 0.43 \\  \hline
 0.80 & 8 $\times$ 10$^{-5}$ & 0/0             & 0 & 1317 & 3316 & 0.93 & 0.40 \\  \hline
 0.80 & 2 $\times$ 10$^{-4}$ & 0/0             & 0 & 1318 & 4033 & 0.93 & 0.33 \\  \hline
 0.80 & 2 $\times$ 10$^{-5}$ & 80/0            & 0 & 1319 & 1381 & 0.93 & 0.96 \\  \hline
 0.80 & 2 $\times$ 10$^{-5}$ & 80/60           & 0 & 1319 & 1180 & 0.93 & 1.11 \\  \hline
 0.80 & 2 $\times$ 10$^{-5}$ & 80/60           & 5 & 1317 & 1271 & 0.93 & 1.04 \\  \hline
 0.80 & 2 $\times$ 10$^{-5}$ & 80/60           & 20 & 1299 & 1301 & 0.91 & 1.0 \\  \hline
 0.90 & 2 $\times$ 10$^{-5}$ & 0/0             & 0 & 1285 & 2619 & 0.91 & 0.49 \\  \hline
 0.90 & 2 $\times$ 10$^{-5}$ & 80/0            & 0 & 1285 & 1179 & 0.91 & 1.09 \\  \hline
\end{tabular}
\end{center}
\caption{Scalar muon selection: number of signal, $\mathrm N_{\rm sig}$, and background, $\mathrm N_{\rm bkg}$, 
events for 2~ab$^{-1}$ of integrated luminosity, selection efficiency, $\epsilon$, and signal over 
background ratio,  $\mathrm N_{\rm sig}/ \mathrm N_{\rm bkg}$, for different probability 
cut, momentum resolution, polarisation and time stamping values.}
\label{tab2}
\end{table}

\newpage
\subsection{Smuon and neutralino mass determination}

The smuon and neutralino masses are extracted from the position of the kinematic edges of the 
muon momentum distribution, a technique first proposed for squarks~\cite{Feng:1993sd}, then 
extensively applied to sleptons~\cite{Martyn:1999tc}:
\begin{eqnarray}
\mathrm {E_{H,L}}=\frac{\sqrt{s}}{4}\left( 1- \frac { m_{\neutralino{1}}^{2} } { m_{\smuon{\pm}_R}^{2} }  \right)
\left( 1 \pm \sqrt{1 - 4  \frac {m_{\smuon{\pm}_R}^{2}} {S}} \right)
\label{formula:eleh}
\end{eqnarray}
The smuon and neutralino masses depend on the beam energy $\sqrt{s}/2$ and the kinematic edges 
$E_{H,L}$ as:
\begin{eqnarray}
m_{\smuon{\pm}_R}=\frac{\sqrt{s}}{2} \left(1-\frac{( E_{H}-E_{L} )^{2}}{( E_{H}+E_{L})^{2}} \right)^{1/2}
\hspace{0.2cm} \mathrm {and} \hspace{0.2cm}
m_{\neutralino{1}}=m_{\smuon{\pm}_R} \left( 1-\frac{ 2 (E_{H}+E_{L})}{\sqrt{s}} \right)^{1/2}
\label{formula:m1m2}
\end{eqnarray}
where $E_{H}$ and $E_{L}$ are the high and low momentum edges of the muon momentum distribution.
This shows that an accurate measurement of the shape of the luminosity spectrum must be achieved
and the value of masses extracted from the momentum spectrum are correlated.
We extract the $\tilde \mu_R$ and \neutralino{1} masses from a 2-par $\chi^2$ fit to the 
reconstructed momentum distribution. The fit is performed with the {\sc Minuit} minimisation 
package~\cite{James:1975dr}. We model the momentum spectrum according to (\ref{formula:eleh}), 
where $\sqrt{s}$ accounts for beamstrahlung and ISR effects, as discussed below. Momentum resolution 
is included through a parametric smearing of the $\rm p_{t}$ distribution for the analysis performed 
at generator level or full tracking for simulated and reconstructed events. 
The fit also accounts for 
the correlations between the $\tilde \mu_R$ and $\neutralino{1}$ masses. 
To investigate the different contributions to the statistical uncertainty on the
smuon and neutralino masses, several fits are performed by changing the 
input conditions.

\subsubsection{Energy spread and ISR}

\begin{figure}[h]
\begin{center}
\begin{tabular}{cc}
 \subfloat[ISR]{\includegraphics[width=7.0cm,clip]{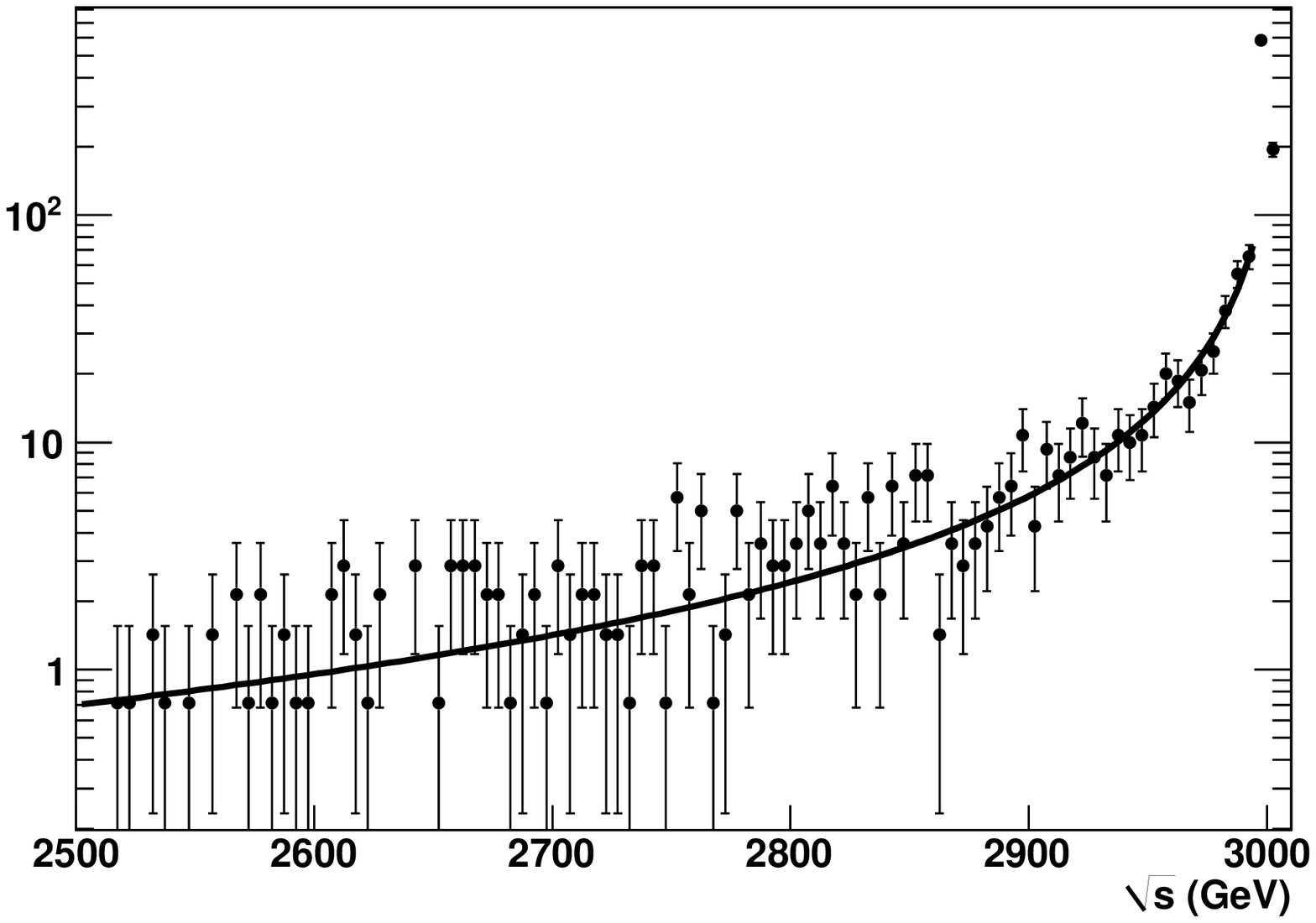}} &
 \subfloat[Beamstrahlung]{\includegraphics[width=7.0cm,clip]{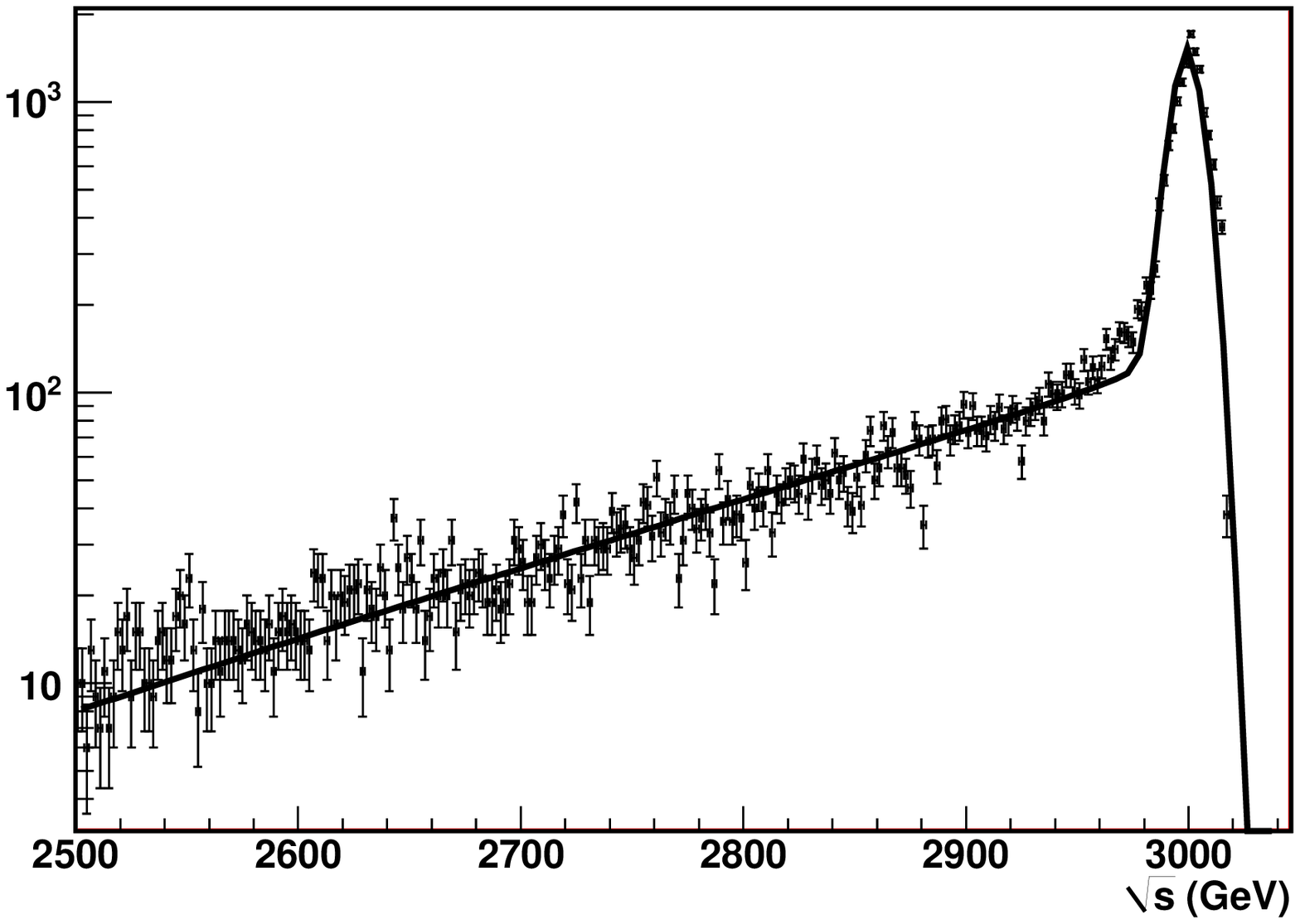}} \\
\end{tabular}
\end{center}
\caption{Centre-of-mass energy distribution including (a) ISR and (b) ISR and beamstrahlung. 
The points represent the simulation 
and the lines the functions used for describing their distribution in the mass fit.}
\label{fig:isr_circe}
\end{figure}

\begin{figure}[h]
\begin{center}
\includegraphics[width=7.5cm,clip]{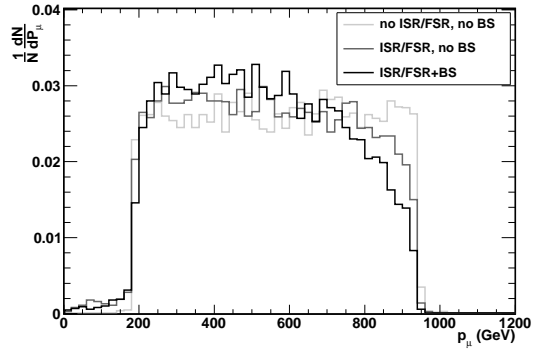} \\
\end{center}
\caption{Signal muon momentum spectrum with no ISR/FSR nor beamstrahlung effects (light grey), 
ISR and FSR only (grey) and also beamstrahlung effects (black) showing the progressive smearing 
of the upper kinematic edge.}
\label{fig:isrbs}
\end{figure}

\begin{figure}[h]
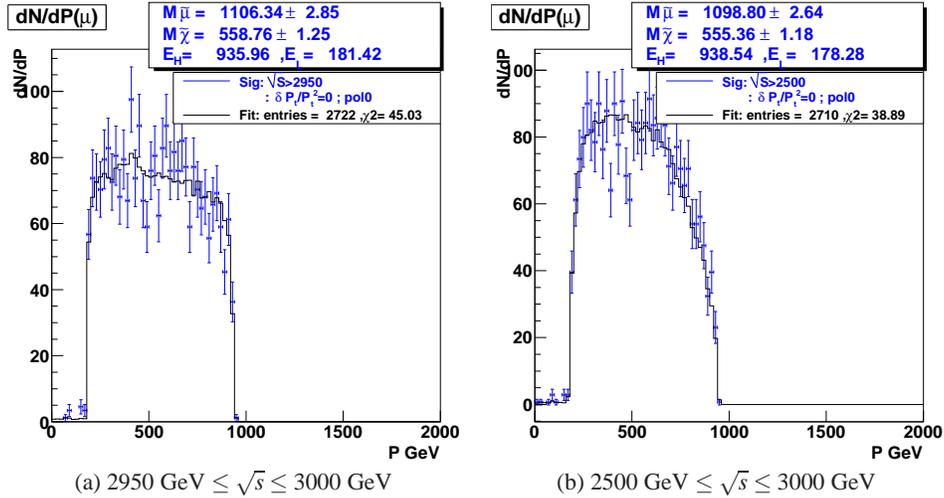

\begin{center}
\begin{tabular}{cc}
 \subfloat[2950~GeV $\le \sqrt{s} \le$ 3000~GeV]
{\includegraphics[width=6.0cm,clip]{MASS_H1LPG205_205_selall_nosmear_pol0_0BX_e2950C50.epsi}} &
 \subfloat[2500~GeV $\le \sqrt{s} \le$ 3000~GeV]
{\includegraphics[width=6.0cm,clip]{MASS_H1LPG205_205_selall_nosmear_pol0_0BX_e2500C50.epsi}} \\
\end{tabular}
\end{center}
  \caption{Fits to the signal muon momentum spectrum for two selections on $\sqrt{s}$.}
 \label{fig:H1AAA}
\end{figure}

We study the contribution of the centre-of-mass energy spread to the statistical accuracy 
of the fit. There are three sources of energy spread: the momentum spread in the linac, which gives 
a $\simeq$7.5~GeV Gaussian smear on $\sqrt{s}$ for the CLIC parameters, beamstrahlung, which 
contributes a long tail and initial state radiation (ISR); the first two are induced by the machine 
and we shall refer to them collectively as ``luminosity spectrum''. We estimate the contribution of 
the luminosity spectrum to the statistical accuracy on the masses and of the knowledge of its shape 
to the mass accuracy and bias. We use the luminosity spectrum obtained from the 
{\sc GuineaPig}~\cite{c:thesis} beam simulation for the 2008 CLIC parameters.
First, we compare the results of the fit for i) events generated without luminosity spectrum spread 
at $\sqrt{s}$ = 3~TeV, ii) events in the main peak of the luminosity spectrum, 2950$< \sqrt{s} <$ 3020~GeV 
and iii) all events with $\sqrt{s} >$ 2500~GeV. In all these cases we apply a loose signal selection and 
we assume no resolution smearing for the muon momentum. Even without the luminosity spectrum 
contribution, the sum of the energies of the colliding electrons extends to energies significantly below 
the nominal $\sqrt{s}$ due to QED effects. We model the ISR spectrum by an approximate solution to the 
Gribov-Lipatov equation, proposed in~\cite{Skrzypek:1990qs}. In the formula we leave free the $\eta$ 
parameter and the fraction of events off the full energy peak. We determine them by a fit to the ISR 
spectrum of {\sc Pythia} signal events (see Figure~\ref{fig:isr_circe}). 
The resulting function is used to fold the ISR contribution in the shape of the muon momentum spectrum 
used in the mass fits. Fig.~\ref{fig:isrbs} shows the effect of ISR and ISR + beamstrahlung on the signal
muon momentum spectrum.
In order to assess the effect of the knowledge of the luminosity spectrum on the mass measurement
accuracy, we consider the luminosity spectrum obtained from {\sc Calypso} for simulated signal 
events and we model it using the parametrisation proposed in~\cite{Ohl:1996fi}. This parametrisation 
has two components: a core, which we assume to be Gaussian, and a tail. We perform a $\chi^2$ to the 
luminosity spectrum with five free parameters: the width of the Gaussian core, two parameters describing 
the tail shape and two normalisation coefficients. The result of the fit is shown in Figure~\ref{fig:isr_circe}. 
Then, we compare the results of the mass fit when we use the fitted parameters of the luminosity 
spectrum parametrisation to those we obtain by varying these by $\pm$15\% of their values in a fully 
correlated way. This change of parameters corresponds to a change of the average $\sqrt{s}$ value by 
$\pm$2$\times$10$^{-3}$. The mass and statistical uncertainty of the smuon change by $\pm$0.8~GeV and 
$\pm$15\%, respectively, and that of the neutralino by $\pm$1.6~GeV and 10\%, respectively. The actual 
accuracy on the determination of the shape of the luminosity spectrum will need to be assessed from a 
detailed study of observables such as the electron acollinearity in Bhabha events~\cite{cliclum}, but 
are expected to be not larger than those assumed here.
Fig.~\ref{fig:H1AAA} (left) shows the fitted muon momentum distribution for events with
2950~GeV $\le \sqrt{s} \le$ 3000~GeV and (right) for events with
2500~GeV $\le \sqrt{s} \le$ 3000~GeV. Results are summarised in Table~\ref{tab:summary}. 
The fitted masses are in agreement with those generated $m_{\smuon{\pm}_R}$ = 1109~GeV and 
$m_{\neutralino{1}}$=554~GeV, within statistical uncertainties.

\subsubsection{Muon photon radiation (FSR)}

A source of resolution loss is photon radiation from muons. At 3 TeV, in about 15\% of the events 
the muon radiates a photon. 
A fit to the muon momentum distribution for signal events applying only a loose selection, 
probability cut=0.5, a cut on the centre-of-mass energy, and without momentum resolution smearing
leads to a small increases of the uncertainty on the neutralino mass, but a shift on the mass value.
(see Table~\ref{tab:summary}).

\subsubsection{Event selection systematics}

The signal selection cut may introduce a bias on muon momentum distribution which 
propagates on the result of the fit to the smuon and neutralino masses. In order to 
study the effect of this cut, we fit the muon momentum distribution for signal events 
with a momentum resolution smearing and two different probability cuts, in the range 
0.8 to 0.99,
For a cut at 0.99 $m_{\smuon{\pm}_R}=1127.6 \pm 3.5$ GeV and $m_{\neutralino{1}}=557.6 \pm 1.7$ GeV.
For a cut at 0.8 $m_{\smuon{\pm}_R}=1104.6 \pm 3.0$ GeV and $m_{\neutralino{1}}=560.0 \pm 1.6$ GeV.
For the events selected with a cut of 0.8 the fitted masses are in agreement with 
those generated, while for the tighter cut at 0.99 results are significantly biased. 
This could be eliminated by applying an efficiency correction which could carry systematic 
uncertainties. Therefore, for this analysis we adopt a selection cut at 0.8, which appears 
safe both in terms of signal-to-background ratio and signal bias.

\subsubsection{Muon momentum resolution}

\begin{figure}[h]
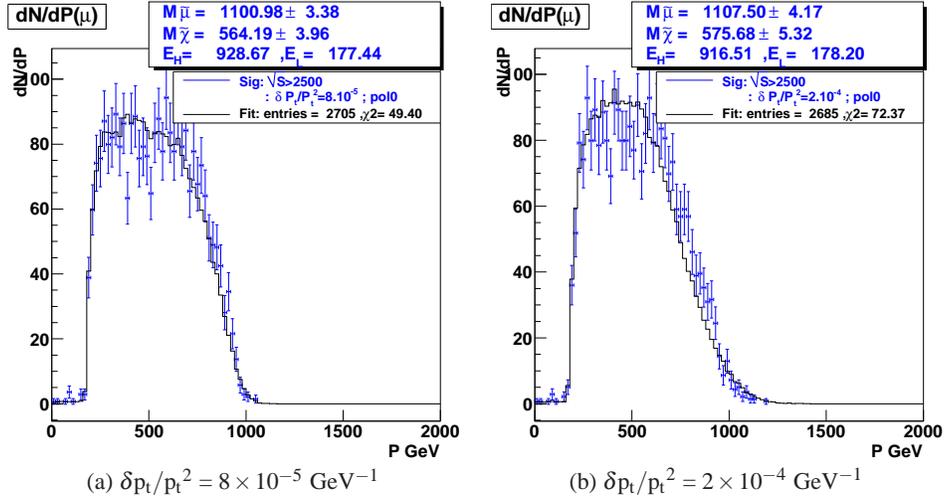

\begin{center}
\begin{tabular}{cc}
\subfloat[$\delta \rm {p_t} / {p_t}^{2}$ = $8 \times 10^{-5}$~GeV$^{-1}$]
{\includegraphics[width=6.0cm,clip]{MASS_H1LPA205_205_selall_smear8e-5_pol0_0BX_e2500C50.epsi}} &
\subfloat[$\delta \rm {p_t} / {p_t}^{2}$ = $2 \times 10^{-4}$~GeV$^{-1}$]
{\includegraphics[width=6.0cm,clip]{MASS_H1LPA205_205_selall_smear2e-4_pol0_0BX_e2500C50.epsi}} \\
\end{tabular}
\end{center}
  \caption{Fits to the signal muon momentum spectrum for momentum smearing of 
(a) $\delta \rm p_{t} / p_{t}^{2}$ = $8 \times 10^{-5}$ and 
(b) $\delta \rm p_{r} / p_{r}^{2}$ = $2 \times 10^{-4}$~GeV$^{-1}$.}
 \label{fig:H1DP2}
\end{figure}

Next, we estimate the contribution of the muon momentum resolution 
on the accuracy of the the masses coming from the fit. In multi-TeV collisions
there is no equivalent of the Higgstrahlung $e^+e^- \to H^0Z^0 \to X \ell^+ \ell^-$, 
($\ell$ = $e$, $\mu$) process, which sets a strict requirement for momentum 
resolution at lower $\sqrt{s}$ values. Reactions such as smuon production in SUSY 
and $H^0 \to \mu^+ \mu^-$ in the SM~\cite{Battaglia:2008aa} can provide useful 
guidance on the track momentum resolution requirements at high energies. We express 
the resolution in terms of $\delta \rm p_{t}/p_{t}^2$, where $\rm p_{t}$ is 
momentum component in the plane normal to the beam axis.    
We perform the mass fit for signal events fulfilling a loose selection and 
2500~GeV $\le \sqrt{s} \le$ 3000~GeV assuming different momentum resolution 
values: 
$\delta \rm {p_t} / {p_t}^{2}$ = 0,
$\delta \rm {p_t} / {p_t}^{2}$ = $2 \times 10^{-5}$~GeV$^{-1}$,
$\delta \rm {p_t} / {p_t}^{2}$ = $4 \times 10^{-5}$~GeV$^{-1}$,
$\delta \rm {p_t} / {p_t}^{2}$ = $6 \times 10^{-5}$~GeV$^{-1}$,  
$\delta \rm {p_t} / {p_t}^{2}$ = $8 \times 10^{-5}$~GeV$^{-1}$
and $\delta \rm {p_t} / {p_t}^{2}$ = $2 \times 10^{-4}$~GeV$^{-1}$.
Fig.~\ref{fig:H1DP1} and \ref{fig:H1DP2} show the fits to the signal muon momentum distribution 
for various momentum resolution values. 
\begin{figure}[h]
\begin{center}
\begin{tabular}{cc}
\subfloat[$\delta \rm {p_t} / {p_t}^{2}$ = $2 \times 10^{-5}$~GeV$^{-1}$]
{\includegraphics[width=6.0cm,clip]{MASS_H1LPA205_205_selall_smear2e-5_pol0_0BX_e2500C50.epsi}} &
\subfloat[$\delta \rm {p_t} / {p_t}^{2}$ = $4 \times 10^{-5}$~GeV$^{-1}$]
{\includegraphics[width=6.0cm,clip]{MASS_H1LPA205_205_selall_smear4e-5_pol0_0BX_e2500C50.epsi}} \\
\end{tabular}
\end{center}
\caption{Fits to the signal muon momentum spectrum for momentum smearing of 
(a) $\delta \rm {p_t} / {p_t}^{2}$ = $2 \times 10^{-5}$~GeV$^{-1}$ and 
(b) $\delta \rm {p_t} / {p_t}^{2}$ = $4 \times 10^{-5}$~GeV$^{-1}$.}
 \label{fig:H1DP1}
\end{figure}
The smuon and neutralino masses are in good agreement with the generated masses. 
The uncertainty on the masses starts being significantly impacted from the momentum 
resolution when $\delta \rm {p_t} / {p_t}^{2}$ is larger than $5 \times 10^{-5}$~GeV$^{-1}$ 
(see Table~\ref{tab:summary}). 

\subsubsection{Background subtraction}

The cross sections for the SM processes which can lead to the same final state as the signal 
are one to two orders of magnitude larger compared to that of the $\tilde \mu_R^+ \tilde \mu_R^-$ 
signal, in absence of beam polarisation. In order to assess the impact of the background on the 
statistical accuracy for the extraction of the $\tilde \mu_R$ and $\tilde \chi^0_1$ masses we 
repeat the analysis to the momentum distribution with both signal and background events. 
\begin{figure}[h!]
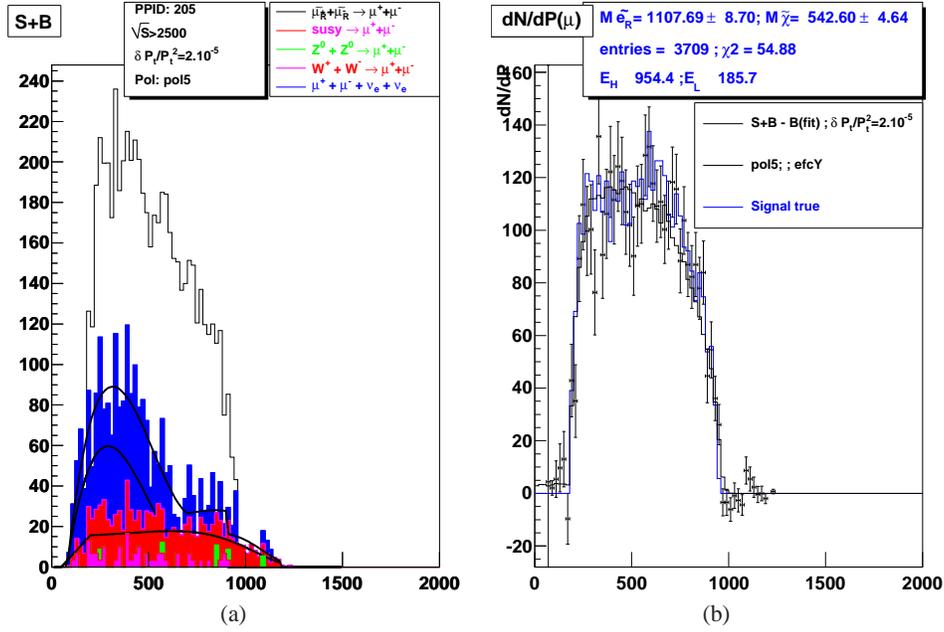

\begin{center}
\begin{tabular}{cc}
\subfloat[]
{\includegraphics[width=6.0cm,clip]{SPB_fsel_H1LPA_205_selall_smear2e-5_pol5_0BX_e2500_C80.epsi}} &
\subfloat[]
{\includegraphics[width=6.0cm,clip]{FITSPB_fsel_H1LPA_205_selall_smear2e-5_pol5_0BX_e2500_C80.epsi}}\\
\end{tabular}
\end{center}
 \caption{(a) Muon momentum spectrum for signal + background events with highlighted the 
different components and the fitted background shape, (b) fit to the 
muon momentum distribution for background-subtracted events. Simulation assumes 80~\% electron
polarisation, momentum resolution $\delta \rm {p_t} / {p_t}^{2}$ = $2 \times 10^{-5}$~GeV$^{-1}$
and selection cut value of 0.8 }
 \label{fig:H1SPB1}
\end{figure}
\begin{figure}[h]
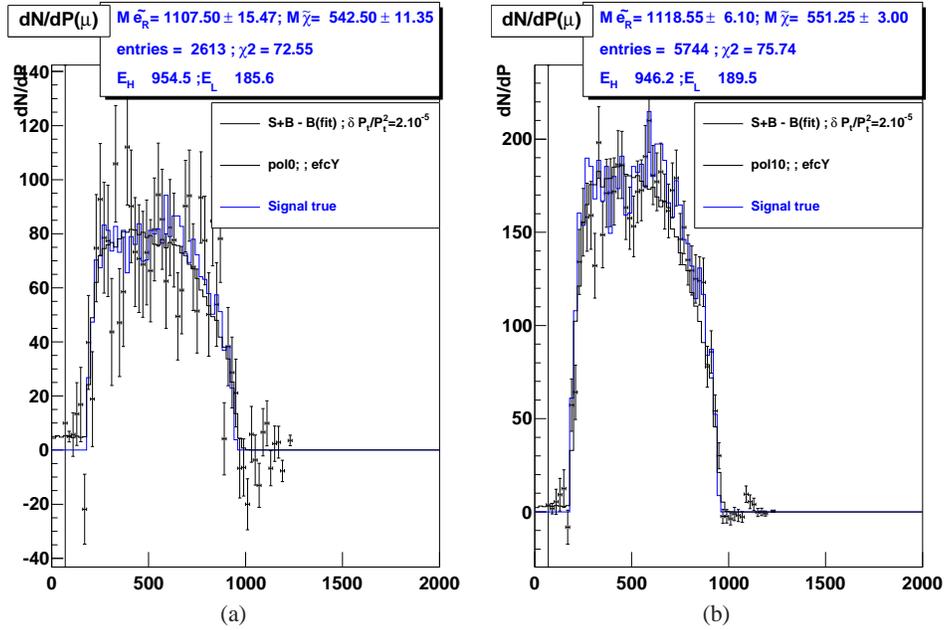

\begin{center}
\begin{tabular}{cc}
\subfloat[]
{\includegraphics[width=6.0cm,clip]{FITSPB_fsel_H1LPA_205_selall_smear2e-5_pol0_0BX_e2500_C80.epsi}} &
\subfloat[]
{\includegraphics[width=6.0cm,clip]{FITSPB_fsel_H1LPA_205_selall_smear2e-5_pol10_0BX_e2500_C80.epsi}} \\
\end{tabular}
\end{center}
 \caption{Fit to the muon momentum distribution for background-subtracted events. Simulation assumes 
(a) no beam polarisation and (b) 80~\% electron and 60~\% positron polarisation, momentum resolution 
$\delta \rm {p_t} / {p_t}^{2}$ = $2 \times 10^{-5}$~GeV$^{-1}$ and selection cut value of 0.8 }
 \label{fig:H1SPB2}
\end{figure}
The $W^+W^-$
background is modelled using an ``ARGUS'' function~\cite{argus} in the range $p_{\mu} >$200~GeV 
and a first order polynomial in the range 100~GeV$< p_{\mu} <$200~GeV. The other backgrounds are 
modelled using a polynomial distribution. These functions are fitted 
on the momentum distribution of background events passing all the selection cuts and used 
to subtract the estimated background contribution from the signal + background momentum 
distribution. After background subtraction the signal distribution is corrected to
take into account the momentum dependent selection efficiency. 
The fit is performed on the background-subtracted momentum spectrum. 
Fig.~\ref{fig:H1SPB1} shows the muon momentum distribution for signal and background
events before (a) and after (b) background subtraction. Events are selected with a probability cut 
of 0.8 and the background is scaled assuming a 80~\% electron beam polarisation. 
Fig.~\ref{fig:H1SPB2} shows the muon momentum distribution for background-subtracted events assuming 
(a) no polarisation and (b) both electron and positron polarisation. 
The polarisation of the electron beam only (option ii)) allows us to improve the measurement of the smuon and 
neutralino masses by 44~\% and 59~\% to a relative statistical accuracy of 0.8\%. Adding positron beam 
polarisation (option iii)) further reduces these uncertainties to 0.6\% and 0.5\%, respectively 
(see Table~\ref{tab:summary}). Background rejection by the use of polarised beams is far 
superior compared to what can be achieved using tighter cuts in absence of polarisation, as shown by a comparison 
of the results obtained with a 0.8 probability cut and electron polarisation to those for a tighter cut at 0.9 
for unpolarised beams in Table~\ref{tab:summary}. A dedicated energy scan of the smuon pair production threshold 
can further improve the measurements of these masses, also reducing their correlation.

\subsubsection{$\gamma\gamma \to {\mathrm{hadrons}}$ Background}

In $e^+e^-$ collisions a high rate of $\gamma\gamma$ collisions arises from photons radiated in the
electro-magnetic interactions.
On average there are about 3.3 $ \gamma\gamma \rightarrow $ hadrons per bunch crossing (BX).
The products of the $\gamma\gamma$ interactions overlap with those from the interactions under 
study. At CLIC, the 312 bunches of a train, separated by 0.5 ns, generate a significant number
of extra particles which are superimposed to the products of the main $e^+e^-$ events 
and degrade the quality of the measurement of its properties~\cite{gghad}.
To estimate the contribution of this background to the uncertainty on the smuon and
neutralino masses, particles from $ \gamma\gamma \rightarrow $ hadrons background are overlayed
on signal and SM events, assuming a detector time stamping capability corresponding to the integration 
of 5~BX and 20~BX. In this analysis the main effect is the change in the efficiency of the signal 
selection. The normalised signal-to-background ratio, S/B, probabilities of the discriminating variables, 
as well as the combined probability {\tt Prob} are computed for a detector resolution:
$\delta \rm p_{t}/p_{t}^{2}$ = 2 $\times$ 10$^{-5}$~GeV$^{-1}$. We find that for the integration 
of 5~BX, the selection efficiency remains virtually unchanged at 0.93, while for 20~BX it becomes 0.91.

\subsubsection{Full Simulation and Reconstruction}

Finally, we repeat the analysis using fully simulated and reconstructed signal events.
The beamstrahlung effects on the the luminosity spectrum are included. 
The simulation is performed using the {\sc Geant-4}-based~\cite{Agostinelli:2002hh} 
{\sc Mokka} program~\cite{MoradeFreitas:2004sq} with the CLIC01-ILD detector geometry, 
which is based on the ILD detector concept being developed for the ILC.
 
\begin{figure}[h]
\begin{center}
\begin{tabular}{cc}
\subfloat[$\delta \rm {p_t} / {p_t}^2$ Resolution from Fully Simulated and Reconstructed Events]
{\includegraphics[width=9.0cm,clip]{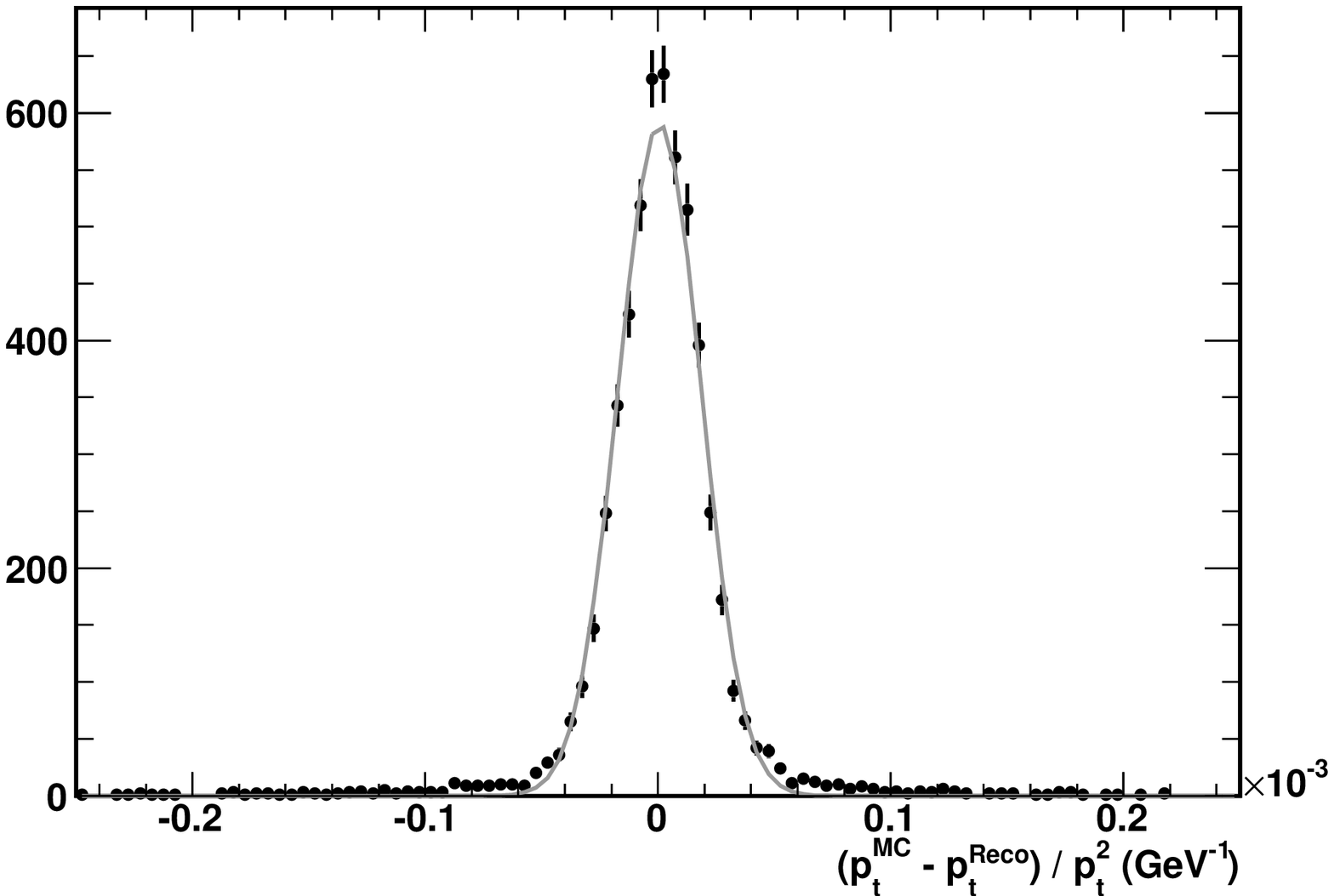}} &
\subfloat[Momentum Spectrum from Fully Simulated and Reconstructed Events]
{\includegraphics[width=6.0cm,clip]{MASS_H1LPA205_205_selall_smear2e-5_pol0_0BX_e2500DST.epsi}}
\end{tabular}
\end{center}
\caption{Validation using fully simulated and reconstructed events for the CLIC01-ILD detector.
(Left) Distribution of the difference between the generated and reconstructed $\rm p_{t}$ 
of muons normalised to the squared $\rm p_{t}$ ($\delta \rm p_{t}/p_{t}^2$), after full simulation 
and reconstruction. The width of the fitted Gaussian curve is 1.8 $\times$ 10$^{-5}$~GeV$^{-1}$.
(Right) Fit to the signal muon momentum spectrum.}
 \label{fig:ptres}
\end{figure}

Events are subsequently reconstructed using the {\sc Marlin} reconstruction 
program~\cite{Gaede:2006pj}. Figure~\ref{fig:ptres} shows the measured momentum resolution 
$\delta \rm p_{t}/p_{t}^2$ obtained for muons in signal events. The masses and accuracies from the fit to 
the fully simulated and reconstructed events, (1118.4$\pm$~3.0)~GeV  and (569.1 $\pm$~1.5)~GeV,
agree with those obtained at generation level with 2$\times$10$^{-5}$~GeV$^{-1}$ momentum smearing 
(see Table~\ref{tab:summary}).

\begin{table}[h]
\caption{Summary of the results of the fits to the smuon and neutralino mass for various assumptions on 
track momentum resolution, beamstrahlung, polarisation and number of bunch crossings integrated in one 
events. The results obtained on signal only (S) at generator level are also compared to those from full 
simulation and reconstruction and signal+background (S+B) fits.}
\begin{center}
\begin{tabular}{|c|c|c|c|c|c|c|}
\hline
$\delta \rm p_{t}/p_{t}^2$      &  $\sqrt{s}>$ & Data   &   Pol           & BX &\multicolumn{2}{c|}{(M$\pm \sigma_M$) (GeV)} \\
($\times 10^{-5}$ GeV$^{-1}$)    &  (GeV)       & Set    &   (e$^-$/e$^+$) &    & $\tilde \mu_R^{\pm}$ & $\tilde \chi_1^0$    \\ \hline
~0.                              &  2950        & S      &   ~0/~0         & ~0   & 1106.3$\pm$~2.9  & 558.8 $\pm$~1.3          \\
~0.                              &  2500        & S      &   ~0/~0         & ~0   & 1098.8$\pm$~2.6  & 555.4 $\pm$~1.2          \\ 
~0.                              &  2500 (ISR only) & S  &   ~0/~0         & ~0   & 1109.2$\pm$~3.2  & 555.4 $\pm$~1.2          \\
\hline
~0.                              &  2500        & S (No FSR Cor) &   ~0/~0  & ~0  & 1095.3$\pm$~3.2  & 
557.7 $\pm$~1.3    \\
\hline
~2.                              &  2500        & S      &   ~0/~0         & ~0  &  1104.6$\pm$~2.9  & 560.0 $\pm$~1.7          \\
~2.                              &  2500        & S (G4+Reco) & ~0/~0      & ~0  &  1107.1$\pm$~2.8  & 560.1 $\pm$~1.5          \\
~4.                              &  2500        & S      &   ~0/~0         & ~0  &  1102.8$\pm$~2.9  & 557.2 $\pm$~2.8          \\
~6.                              &  2500        & S      &   ~0/~0         & ~0 &  1098.8$\pm$~3.1  & 559.1 $\pm$~3.6          \\
~8.                              &  2500        & S      &   ~0/~0         & ~0  &  1101.0$\pm$~3.4  & 564.2 $\pm$~4.0    \\
20.                              &  2500        & S      &   ~0/~0         & ~0  &  1107.5$\pm$~4.2  & 575.7 $\pm$~5.3   \\ \hline
~2.                              &  2500        & S+B (0.8) &   ~0/~0      & ~0  &  1107.5$\pm$15.5  & 542.5 $\pm$~11.3  \\
~2.                              &  2500        & S+B (0.9) &   ~0/~0      & ~0  &  1107.5$\pm$14.4  & 551.2 $\pm$~12.0  \\
~2.                              &  2500        & S+B (0.8) &   80/~0      & ~0  &  1107.7$\pm$~8.7  & 542.6 $\pm$~4.6   \\
~2.                              &  2500        & S+B (0.8) &   80/60      & ~0  &  1118.5$\pm$~6.1  & 551.3 $\pm$~3.0  \\ 
\hline 
~2.                              &  2500        & S+B (0.8) &   80/60      & ~5  &  1105.7$\pm$~6.3  & 549.4 $\pm$~3.9  \\ 
~2.                              &  2500        & S+B (0.8) &   80/60      & 20  &  1113.2$\pm$~6.8  & 550.3 $\pm$~3.4  \\ 
\hline
\end{tabular}
\end{center}
\label{tab:summary}
\end{table}

\subsection{Summary}

This study allows us to draw some conclusions on the potential of a 3~TeV 
CLIC collider in SUSY spectroscopic measurements and some of the requirements on the 
detector and the beams. Because of the tiny production cross section in the chosen high-mass 
scenario, background subtraction is the dominant source of statistical uncertainty.
Electron beam polarisation at $\simeq$~80~\% gives an equivalent luminosity gain of a factor 
of six and is essential to recover precision. Positron polarisation is desirable, since it  
gives an additional gain of a factor of two in equivalent luminosity and it also allows us to 
disentangle the contributions of $\tilde \mu_L$ and $\tilde \mu_R$.  Smuon and neutralino masses 
of 1108.8~GeV and 554.3~GeV, respectively can be extracted from the muon kinematics, in events 
with two oppositely charged muons and missing energy, with a relative statistical accuracy 
$\sim$~0.5~\% with 2~ab$^{-1}$ of integrated luminosity and both beams polarised. 
In addition, the signal production cross section of 0.7 fb can be determined 
with a relative statistical uncertainty of 2.0~\%. 

Since a major source of smearing of the kinematic edges of the muon momentum spectrum 
is beamstrahlung and ISR, the track momentum resolution does not appear to be critical for 
the measurement of the smuon mass, as long as a resolution 
$\delta \rm p_{t}/p_{t}^2 \le 5 \times 10^{-5}$~GeV$^{-1}$ can be achieved, though it remains
important for the neutralino mass.  It is important to have a good control of the luminosity 
spectrum and desirable to limit the beamstrahlung not significantly beyond that corresponding 
to the 2008 CLIC parameters. 

Finally, the effect of the overlay of $\gamma \gamma \to {\mathrm{hadrons}}$ events from machine-induced 
background does not lead to any significant degradation of the signal selection efficiency for a detector 
with time stamping capability of 10~ns.

%
\section{Acknowledgements}

We are grateful to Daniel~Schulte for making the luminosity spectrum and 
generated $\gamma \gamma \to \mathrm{hadrons}$ events available and to Dieter Schlatter
for his careful reading of this note.


\begin{thebibliography}{100}
\bibitem{c:thesis} 
  D.~Schulte, TESLA Note 97-08.\\
  The results of the beam simulation have been provided by Daniel Schulte.
\end{thebibliography}

%


\end{document}